\documentclass[aps,pra,floatfix,twocolumn,superscriptaddress,showpacs,10pt]{revtex4-1}
\usepackage{amssymb, amsmath,color,mciteplus,graphicx,subfigure}
\usepackage{calc,epsfig,epstopdf,color,mciteplus,bm,mathrsfs}
\usepackage{times}
\usepackage{float}
\usepackage{lipsum}
\usepackage{xcolor}

\begin{document}

\title{Time-optimal universal quantum gates on superconducting circuits}

\author{Ze Li}

\author{Ming-Jie Liang}
\affiliation{Key Laboratory of Atomic and Subatomic Structure and Quantum Control (Ministry of Education),  Guangdong Basic Research Center of Excellence for Structure and Fundamental Interactions of Matter, and School of Physics, South China Normal University, Guangzhou 510006, China}

\author{Zheng-Yuan Xue}\email{zyxue83@163.com}
\affiliation{Key Laboratory of Atomic and Subatomic Structure and Quantum Control (Ministry of Education),  Guangdong Basic Research Center of Excellence for Structure and Fundamental Interactions of Matter, and School of Physics, South China Normal University, Guangzhou 510006, China}

\affiliation{Guangdong Provincial Key Laboratory of Quantum Engineering and Quantum Materials, Guangdong-Hong Kong Joint Laboratory of Quantum Matter,  and Frontier Research Institute for Physics,\\ South China Normal University, Guangzhou 510006, China}

\date{\today}

\begin{abstract}
Decoherence is inevitable when manipulating quantum systems. It  decreases the quality of quantum manipulations  and thus is one of the main obstacles for large-scale quantum computation, where high-fidelity quantum gates are needed. Generally, the longer a gate operation is, the more decoherence-induced gate infidelity will be. Therefore, how to shorten the gate  time becomes an urgent problem to be solved. To this end, time-optimal control  based on solving the quantum brachistochrone  equation is a straightforward solution. Here, based on time-optimal control,  we propose a scheme to realize universal quantum gates  on superconducting qubits   in a two-dimensional  square  lattice configuration, and the two-qubit gate fidelity approaches $99.9\%$. Meanwhile, we can further accelerate the {\it Z}-axis gate considerably  by adjusting the detuning of the external driving. Finally, in order to reduce the influence of the dephasing error, decoherence-free subspace encoding is also incorporated  in our physical implementation. Therefore, we present a  fast quantum scheme which is promising for  large-scale quantum computation.
\end{abstract}

\maketitle

\section{Introduction}
Due to the intrinsic superposition nature,  quantum computation can   deal with problems that are hard for classical computers. Recently, quantum computation has been proposed to be implemented on  various quantum platforms  \cite{1,2,3,4}, among which, the superconducting quantum circuits system is one of the most promising candidates \cite{5,6,7,8}. However, besides  the  existence of  operational errors, a quantum system will inevitably couple to its surrounding environment,  and thus  lead to an increase in the distortion of quantum states or operations. Therefore, how to achieve high fidelity quantum gates on quantum systems is an urgent problem to be solved.

In the presence of noise, high-quality  quantum control can be realized by the fastest possible evolution. Therefore, finding a shorter gate evolution path to shorten the gate time has become an effective means to achieve high-fidelity quantum gates \cite{liang}. Time-optimal control (TOC)  based on solving the quantum brachistochrone  equation (QBE) \cite{QBE} is an effective method  to shorten the evolution time \cite{10}. Recently,  TOC-based schemes for unitary operations have been proposed \cite{10, 11, 12, ch, 13, 14, 15, 16} and experimentally demonstrated \cite{e1, e2, e3, 17, 18, 19}, where the  time needed for specific quantum gates  has been reduced significantly. However, universal quantum control with an analytical solution can only be possible for specific cases \cite{11}.

Here, based on TOC,  we propose a scheme to realize universal quantum gates  on superconducting   transmon qubits,  arranged    in a two-dimensional  (2D) square  lattice configuration,  which can support  large-scale universal quantum computation. In our scheme, by time-dependent  modulation of a superconducting qubit, we can achieve tunable coupling between two transmon qubits \cite{tune0, tune1, tune2}, which can be readily used to induce target quantum gates in only one step.  Meanwhile, we can further shorten the time for  $Z$-axis rotations by adjusting the time-independent driving detuning. Note that, in the previous work \cite{13}, $Z$-axis rotational gates can only be achieved by two sequential steps,  i.e., by a combination of $X$-and $Y$-axis rotations,  thus leading to longer gate times.  Furthermore,  to eliminate the effect of collective dephasing, which is an  important factor affecting the  gate fidelity, different from  the previous work \cite{13} ,  decoherence-free subspace (DFS) encoding  \cite{dfs1, dfs2, dfs3} has been incorporated  and the robustness of our gates with respect to the decoherence is verified. Therefore, our work realized high-fidelity universal quantum gate on superconducting circuits, which is a  promising scheme for future large-scale quantum computation.

\section{Quantum gates via TOC}
As it is well-known,quantum computation can be implemented based on two-level quantum systems, i.e., using qubit systems. Therefore, we begin with presenting a general method of constructing quantum gates via TOC, based on a qubit system,  denoted by $\{|0\rangle=\left ( 1,0 \right ) ^{\dagger }, |1\rangle=\left ( 0,1 \right ) ^{\dagger } \}$. Assuming $\hbar=1$, the  general light-matter interaction can be written as
\begin{eqnarray}\label{YS}
\mathcal{H} (t)& = & \frac{1}{2}\left(\begin{array}{cc}
\delta(t) & \Omega(t) e^{-i \phi(t)} \\
\Omega(t) e^{i \phi(t)} & -\delta(t)
\end{array}\right),
\end{eqnarray}
where $\Omega(t)$ and $\phi(t)$ are the amplitude and the  phase of the driving field, respectively, and $\delta(t)$ is the   detuning between the qubit frequency  and the  driving field frequency. Choosing two mutually orthogonal evolution states, $|\Psi_{\pm}(t)\rangle$, that satisfy the time-dependent Schr\"{o}dinger equation of Hamiltonian in Eq. (\ref{YS}), the evolution operator can be written as
 \begin{eqnarray}
U(t)&=&\mathbf{T} e^{i \int \mathcal{H}(t) \mathrm{d} t}\notag\\
&=&\left|\Psi_{+}(t)\right\rangle\left\langle\Psi_{+}(0)\right| +\left|\Psi_{-}(t)\right\rangle\left\langle\Psi_{-}(0)\right|,
 \end{eqnarray}
where $\mathbf{T}$ is the time-ordering operator.

In order to construct a particular evolution operator, we need to define a set of auxiliary basis vectors, $\left|\psi_{\pm}(t)\right\rangle=e^{-i \gamma_{\pm}(t)}\left|\Psi_{\pm}(t)\right\rangle$, with $\gamma_{\pm}(0)=0$ and $\gamma_{+}(t)=-\gamma_{-}(t)$, which satisfy the  boundary condition of $|\psi_\pm(\tau)\rangle=|\psi_\pm(0)\rangle=|\Psi_\pm(0)\rangle$. We select a pair of dressed states,
\begin{eqnarray}
\left|\psi_{+}(t)\right\rangle & = & \cos \frac{\chi(t)}{2}|0\rangle+\sin \frac{\chi(t)}{2} e^{\mathrm{i} \xi(t)}|1\rangle, \notag\\
\left|\psi_{-}(t)\right\rangle & = & \sin \frac{\chi(t)}{2} e^{-\mathrm{i} \xi(t)}|0\rangle-\cos \frac{\chi(t)}{2}|1\rangle,
\end{eqnarray}
as a set of auxiliary basis vectors, which are the eigenstates of the Lewis-Riesenfeld  invariant \cite{22} 
\begin{eqnarray}
\label{LR}
I(t) & = & \frac{\mu}{2}\left(\begin{array}{cc}
\cos \chi(t) & \sin \chi(t) e^{-i \xi(t)} \\
\sin \chi(t) e^{i \xi(t)} & -\cos \chi(t)
\end{array}\right)
\end{eqnarray}
of Eq. (\ref{YS}), where $\mu$ is an arbitrary constant. The auxiliary basis vectors $\left|\psi_{\pm}(t)\right\rangle$  on the Bloch sphere show their geometric evolutionary details through  the parameters $ \xi(t) $ and $\chi(t)$.

Then, by solving the dynamic invariant equation of $i \partial I(t) / \partial t-[\mathcal{H}(t), I(t)]=0$, the parameters $\xi(t)$ and $\chi(t)\}$ of $\left|\psi_{\pm}(t)\right\rangle$  are decided by the parameters $\Omega(t)$, $ \phi(t)$, and $\delta(t)$ of  Eq. (\ref{YS}) as \cite{22,23,24}
\begin{eqnarray}\label{XZ1}
\dot{\chi}(t)&=&\Omega(t) \sin [\phi(t)-\xi(t)], \notag\\
\dot{\xi}(t)&=&\delta(t)-\Omega(t) \cot \chi(t) \cos [\phi(t)-\xi(t)].
\end{eqnarray}
After an evolution time $\tau$, by solving the Schr\"{o}dinger equation $\mathcal{H} (t)\left | \Psi_{\pm}(t)  \right \rangle = i\hbar \frac{\partial }{\partial t} \left | \Psi_{\pm}(t)  \right \rangle$, we can obtain the overall phase as
\begin{gather}
\gamma(\tau)=\int^{\tau}_{0}\frac{2 \dot{\xi}(t) \sin ^{2}\left[\chi(t)/2\right] -\delta(t)}
{2 \cos \chi(t)}dt.
\label{XZ2}
\end{gather}
Equations (\ref{XZ1}) and  (\ref{XZ2}) show the correspondence between the Hamiltonian parameters $\{ \Omega(t), \phi(t), \delta(t)\}$ and the $U(t)$ parameters $\{ \gamma(t), \xi(t), \chi(t)\}$.
By setting $\chi(t)$ and $\delta(t)$ to be constants, for less experimental demanding, Eq. (\ref{XZ1}) reduces to
\begin{eqnarray}
\xi(t)&=& \phi(t)  -\pi, \notag\\
\cot \chi &=&[\dot{\phi}(t)-\delta ]/\Omega(t) ,\label{XZ4}
\end{eqnarray}
as $\Omega (t)$ can not be zero. In this way, we can get a general evolution operator of the process as
\begin{eqnarray}
\label{YH1}
U(\tau)&&=\cos \gamma^{\prime}\left(\begin{array}{cc}
 e^{-\mathrm{i} \phi ^{-}} &  0 \\
 0 &  e^{\mathrm{i} \phi ^{-}}
\end{array}\right)\notag\\
&&+\mathrm{i} \sin \gamma^{\prime}
\left(\begin{array}{cc}
 \cos \chi e^{-\mathrm{i} \phi ^{-}} &  \sin \chi e^{-\mathrm{i} (\phi ^{+}-\pi)} \\
 \sin \chi e^{\mathrm{i} (\phi ^{+}-\pi)} & - \cos \chi e^{\mathrm{i} \phi ^{-}}
\end{array}\right),
\end{eqnarray}
where $\gamma{'}=\gamma(\tau)+\phi ^{-}$ and $\phi ^{\pm } =[\phi (\tau )\pm\phi (0 )]/2 $.
We can construct arbitrary target quantum gates  by controlling the parameters $\gamma$ and $\chi$ via the external coupling strength $\Omega(t)$ and the phase $\phi(t)$. That is, our purpose is to find the general relations between Hamiltonian parameters,  see Eq. (\ref{XZ1}), so that  quantum gates in Eq. (\ref{YH1}) can be obtained. 

Considering  experimental implementations, the interaction term in Eq. (\ref{YS}),  $\mathcal{H}_{c}(t)= \Omega(t)\left[\cos \phi(t) \sigma_{x}+\sin \phi(t) \sigma_{y}\right]/2$ with $\sigma_{x, y, z}$ being the Pauli matrices, needs to satisfy the following two conditions. First, the coupling strength $\Omega(t)$ is adjusted within a restricted range, so there is an upper bound on $\Omega(t)$. $f_1(\mathcal{H}_{c}(t))=[\text{Tr}(\mathcal{H}_{c}(t))^{2} -\Omega(t) ^{2}/2]/2=0$ needs to be satisfied. Then, the form of the interaction term $\mathcal{H}_{c}(t)$ is usually not arbitrary and an independent $\sigma_z$ operator cannot be achieved, so it is necessary to satisfy $f_2(\mathcal{H}_{c}(t)) =\text{Tr}(\mathcal{H}_{c}(t) \sigma_{z} )=0$. Considering   these two conditions, based on the QBE of
\begin{eqnarray}
\frac{d F}{d t}=-i[\mathcal{H}(t), F],
\end{eqnarray}
where
\begin{eqnarray}\label{Line}
F=\frac{\partial\left(\sum_{j=1,2} \lambda_{j} f_{j}(\mathcal{H}_c(t))\right)}{\partial \mathcal{H}_c(t)}
=\lambda_{1}\mathcal{H}_c(t)+\lambda_{2} \sigma_{z},
\end{eqnarray}
and the Lagrange multiplier $\lambda_{j}$ is defined as $\lambda_{1}=1/\Omega(t)$ and $\lambda_{2}=-c/2$, where $c$ is a constant, we can obtain  $\phi(t)=\phi(0)+\phi^{\prime}(t)$ and $\phi^{\prime}(t)=\int_{0}^{t}\left[ c\Omega({t^{\prime}}) +\delta\right] \mathrm{d} t^{\prime}$. 
For a specific gate, $\phi^-$ is set, and the gate time $\tau$ is determined by the equation of $\int^{\tau}_{0} \dot{\phi}(t) dt=2\phi^-$. To get the shortest operation time $\tau$, $\dot{\phi}(t)$ should be a constant. Then, by solving Eq. (\ref{XZ4}), $\Omega(t)$ is a constant too, which is set as its maximum $\Omega(t)=[\Omega(t)]_{max}=\Omega$ according to specific systems. That is, to implement a fast quantum gate, the coupling strength, detuning, and phase need to satisfy the conditions   $\dot{\Omega}
(t)=0$, $\dot{\delta},
(t)=0$ and $\dot{\phi}(t)=\eta$, respectively, where $\eta$ is constant.

Therefore, based on TOC, the evolution operator in Eq. (\ref{YH1}) can be rewritten as
\begin{eqnarray}
\label{YH2}
U(\tau)&=&\left(\begin{array}{cc}
 e^{-\mathrm{i} \phi^-} &  0 \\
 0 &  e^{\mathrm{i} \phi^-}
\end{array}\right) U_{g},
\end{eqnarray}
\begin{eqnarray}
\label{YH3}
U_{g}=\cos \gamma^{\prime}
+\mathrm{i} \sin \gamma^{\prime}
\left(\begin{array}{cc}
 \cos \chi  &  -\sin \chi e^{-\mathrm{i} \phi(0)} \\
 -\sin \chi e^{\mathrm{i} \phi(0)} & - \cos \chi
\end{array}\right),
\end{eqnarray}
where
\begin{gather}
\cot \chi =\frac{\eta - \delta }{\Omega},\notag\\
\gamma' =\frac{\eta \tau }{2 }+ \int^{\tau}_{0}\frac{2 \eta \sin ^{2}\left[\chi/2\right] -\delta}
{2 \cos \chi}dt,
\label{XZ3}\end{gather}
and $U_{g}$ can be used to obtain arbitrary single-qubit gates.  
Note that, the operation in  Eq. (\ref{YH3}) can be implemented in a single-step way, instead of being constructed by sequential gates in conventional cases. Besides, in the above example, an arbitrary single-qubit gate $U_{g}$ can be obtained by simple square pulses. However, we want to emphasize that the parameters of the driving field are only required to meet the condition in Eq. (\ref{XZ4}); i.e., $\Omega(t)$ can be time-dependent in general, and thus further pulse shape is also allowed.

Now, we calculated the gate operation time $\tau$  by solving   Eqs. (\ref{YH3}) and  (\ref{XZ3}). We here take the universal gate set of \{$H$, $S$,  $T$\} as a typical example.  For $S$ and $T$ gates, setting $\gamma^{\prime}=\pi$, $\phi_S^-=-3\pi/4$, and $\phi_T^-=-7\pi/8$, we can obtain the $S$ gate and the $T$ gate, respectively. The gate times are
\begin{eqnarray}
&&\tau_S =\frac{\pi}{2(\Omega^2+\delta^2)}\left( \sqrt{16 \delta^2 + 7 \Omega^2 } - 3\delta \right), \notag\\
&&\tau_T =\frac{\pi}{4(\Omega^2+\delta^2)}\left( \sqrt{64 \delta^2 + 15 \Omega^2 } - 7\delta \right),
\end{eqnarray}
where $\delta$ can be adjusted to further accelerate the  gates, due to the extra freedom on $\chi$. 

For the $H$ gate, $\chi_H=\pi/4$, so that  $\eta_H-\delta_H=\Omega_H$. As  $\gamma'_H=\pi/2$, from Eq. (\ref{XZ3}), we get   $\tau_H=\pi/ (\sqrt{2}\Omega_H).$ 
When $\phi_0=(2n+1)\pi$, with $n$ being an arbitrary integer, the evolution operator in Eq. (\ref{YH2}) reduces to
\begin{eqnarray}
U_H(\tau) & = & \left(\begin{array}{cc}
e^{i(\pi- \eta \tau)/2} & 0 \\
0 & e^{ i  (\pi +\eta \tau)/2}
\end{array}\right)
\left(\begin{array}{cc}
1 & 1 \\
1 & -1
\end{array}\right).
\end{eqnarray}
When we set $\eta\tau =2\pi$, we obtain an H  gate, and $\tau_H$ is only determined by $\Omega=\eta-\delta$;  changing one of them will lead to the change of the other one, and thus no acceleration of gate time can be obtained. 

Note that, the implementation of our proposal with  superconducting qubits is straightforward, as the Hamiltonian in Eq. (\ref{YS}) is readily realizable experimentally \cite{tune0} by directly applying a microwave drive to a qubit. Besides the tunable coupling of two qubits   \cite{tune0, tune1, tune2} can be used to construct two-qubit gates. However, this direct implementation is  limited by the weak anharmonicity of transmon qubits and their crosstalk-induced {\it Z} error.

\begin{figure}[tbp]
	\includegraphics[width= \columnwidth]{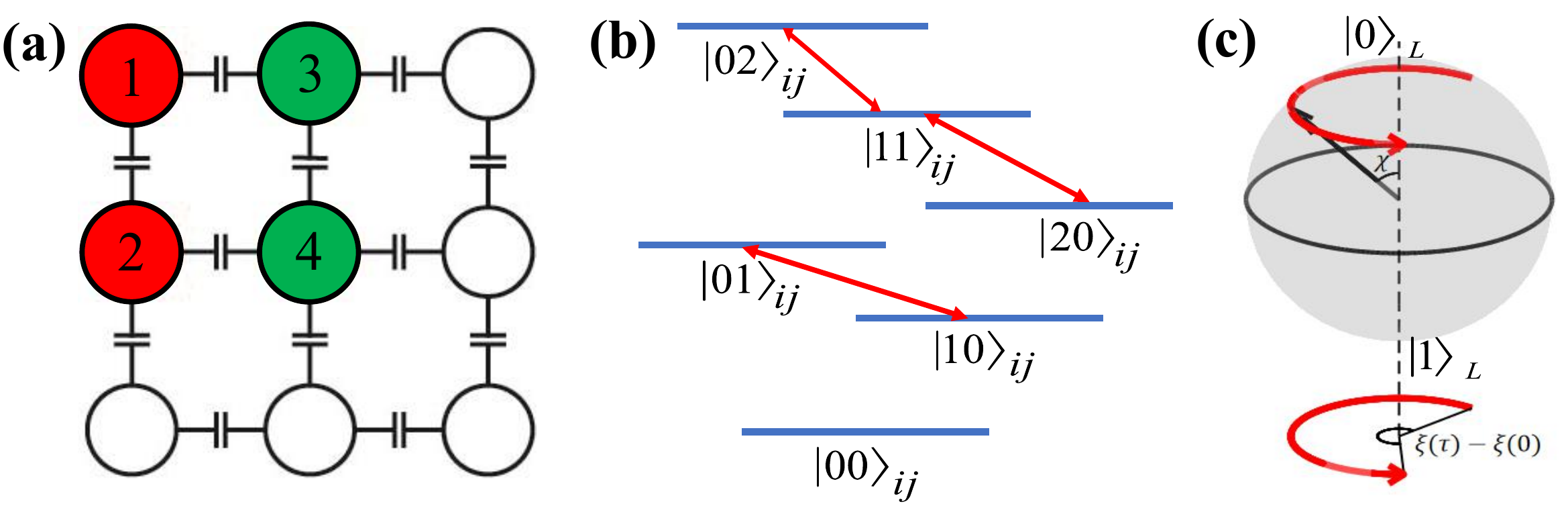}
	\caption{Illustration of our scheme. (a) A scalable 2D square  lattice consists of transmon qubits, where adjacent qubits are capacitively coupled. Two physical qubits of the same color encoded as a DFS logical qubit. (b) The energy levels of  two adjacent  coupled qubits, $T_i$ and $T_j$, where different excitation subspaces can be used to implement different quantum gates. (c) Illustration of the evolution path of the TOC-based scheme (red line) on the Bloch sphere, where $\chi$ is the angle between the direction of the auxiliary basis vector and the vertical axis, and $\xi(\tau)-\xi(0)$ is the horizontal angle shift of the auxiliary basis vector at a specific time $\tau$.}
\label{fig1}
\end{figure}

\section{Physical implementation with encoding}
Here, to further decrease the {\it Z} errors, induced by the qubit-crosstalk and the dephasing effect of physical qubits,  we incorporate the DFS encoding  in our scheme. We consider the implementation of  the TOC  scheme with encoding based on the 2D square lattice consisting of superconducting transmon qubits, as shown in Fig. \ref{fig1}(a), where a transmon  serves  as a physical qubit. Labeling two adjacent transmon qubits to be qubits $T_1$ and $T_2$ as shown in Fig. \ref{fig1}(a),  the logical qubits  can be encoded in their single-excitation subspace, i.e., $S_1=\text{Span}\{|0\rangle_{L}=|10\rangle_{12}, |1\rangle_{L}=|01\rangle_{12}\}$.
By this encoding, the logical qubit   can resist  the collective {\it Z} error of the physical  qubits. Besides,  this single-excitation subspace encoding can effectively suppress the leakage error in the single physical qubit case, due to the weak anharmonicity of transmon qubits, as the transition between different excitation subspaces are energetically suppressed. 

\subsection{Single-logical-qubit gates via TOC}
As the coupling strength between two adjacent transmons is usually fixed, to control single-logical-qubit units and two-logical-qubit units independently and construct the targeted quantum gates exactly, tunable interactions between any two transmon qubits should be achieved. 
For two adjacent transmon qubits $T_i$ and $T_j$, the interaction Hamiltonian is
\begin{eqnarray}
\mathcal{H}^{0}_{i j}&=&\sum_{k=i,j}\left[\omega_{k}|1\rangle_{k}\left\langle1\left|+\left(2 \omega_{k}-\alpha_{k}\right)\right|2\right\rangle_{k}\langle2|\right] \notag\\
&+& g_{i j}(|10\rangle_{i j}\langle 01|+\sqrt{2}| 11\rangle_{i j}\langle 02|\notag\\
&+&\sqrt{2}|20\rangle_{i j}\langle 11| +\text{H.c.}),
\end{eqnarray}
where $\omega_{i,j}$ and $\alpha_{i,j}$ are the frequency and the anharmonicity of the $i$the and $j$th transmon qubit $T_i$ and $T_j$, respectively, and $|\text{CD}\rangle_{i j}=|\text{C}\rangle_{i} \otimes |\text{D}\rangle_{j}$.
To achieve  tunable coupling  between $T_i$ and $T_j$, we added a frequency modulation in the form of $\omega_j=\omega_{j0}+ \epsilon_j \cos[\nu_j t+\phi_j(t)]$ for qubit $T_{j}$, with the driving frequency and the phase being $\nu_j$  and $\phi_j(t)$, respectively. Meanwhile, the frequency of $T_{i}$ is fixed, which is written as $\omega_{i}=\omega_{i0}$ for the same layout as $\omega_{j}$. Moving into the interaction picture with respect to
\begin{equation}
    U_{ij}^{I}=U_{i}^{I}\times U_{j}^{I},
\end{equation}
with
\begin{equation}
U_{i}^{I}=\exp[-\text{i}(\omega_{i0} b_{i}^{+}b_{i}-\frac{\alpha_i}{2} b_{i}^{+}b_{i}^{+}b_{i}b_{i}) t ],
\end{equation}
\begin{eqnarray}
U_{j}^{I}&=&\exp\{-\text{i}[\omega_{j0} t+\Gamma_j \sin(v_j t+\phi_j(t))] b_{j}^{+}b_{j}\notag\\
&&-\text{i}\frac{\alpha_j}{2} b_{j}^{+}b_{j}^{+}b_{j}b_{j}t \},
\end{eqnarray}
with $b_{i,j}= (| 0\rangle_{i,j}\langle 1|+\sqrt{2} |1\rangle_{i,j}\langle2| )$, $\Gamma_j=\epsilon_j/[\nu_j+\dot{\phi}_j(t)]$, and then using the Jacobi-Anger identity, $\exp(-i\Gamma \sin\theta)=\sum_{n} J_{n} (\Gamma ) \exp(-in\theta)$, where $J_n$ is the $n$th Bessel function, the transformed Hamiltonian can be written as
\begin{eqnarray}
    \label{HT1t}
    \mathcal{H}^{I}_{i j}&=&g_{i j} e^{i \Delta_{ij}} \mathcal{K}_{j}\{|10\rangle_{i j}\langle 01|
    +\sqrt{2}e^{i \alpha_j t}| 11\rangle_{i j}\langle 02|\notag\\
    &&+\sqrt{2}e^{-i \alpha_i t}| 20\rangle_{i j}\langle 11|\}+\text {H.c.} ,
\end{eqnarray}
where $\Delta_{ij}=-\Delta_{ji}=\omega_{i0}-\omega_{j0}$ is the frequency difference between $T_i$ and $T_j$, and $\mathcal{K}_{j}=\sum_{n  =  -\infty}^{+\infty}  J_{n}\left(\Gamma_{j}\right)\exp[-i n(\nu_{j}t+\phi_{j})]$. The energy spectrum is shown in Fig. \ref{fig1}(b), and adjacent levels within a certain excitation subspace can be used to realize different quantum gates. In addition, $\Gamma_j$ can be tuned  to achieve adjustable coupling between qubits $T_i$ and $T_j$, and thus  we can select  appropriate parameters of the modulation field to construct target quantum gates.

In order to obtain a two-level system Hamiltonian in the form of Eq. (\ref{YS}) for constructing universal quantum gates, it is natural to go into the rotating frame with respect to
\begin{equation}
U_{A} =\exp \left[i\frac{\delta}{2} t (|0\rangle_{L}\langle 0|-|1\rangle_{L}\langle 1|)\right],
\end{equation}
and the transformed Hamiltonian of Eq. (\ref{HT1t}) can be written as
\begin{eqnarray} \label{KK}
\mathcal{H}_{1 2}^{\prime}
&=& \frac{\delta}{2} (|0\rangle_{L}\langle 0|-|1\rangle_{L}\langle 1|), \notag\\
&+& g_{1 2}\{\mathcal{K}_{2}e^{i \Delta_{1 2}t}( e^{-i \delta t}|0\rangle_{L}\langle 1|
    +\sqrt{2}e^{i \alpha_j t}| 11\rangle_{1 2}\langle 02|\notag\\
    &+&\sqrt{2}e^{-i \alpha_i t}| 20\rangle_{1 2}\langle 11| ) +\text { H.c.}\},
\end{eqnarray}
Choose the modulating  frequency to meet $\nu_2=\Delta_{12}-\delta$ in Eq. (\ref{KK}), and after the rotational wave approximation, we obtain the Hamiltonian in the logical basis $S_1$ as
\begin{eqnarray}
\mathcal{H}^{\text{eff}}_{12}=\frac{1}{2}\left(\begin{array}{cc}
\delta& \Omega e^{-i \phi_2} \\
\Omega e^{i \phi_2} & -\delta
\end{array}\right),
\end{eqnarray}
where $\Omega=2g_{12}J_1(\Gamma_2)$.
By adjusting the pulse parameters $\epsilon_2$, $\nu_2$, and $\phi_2$, we can find a path that accords with the TOC-based scheme. Therefore, according to the general theory in the last section, we can use the TOC-based scheme to construct  arbitrary single-logical-qubit quantum gates. We set different parameters of the physical qubits for $the H$, $S$ and, $T$ gates.
The $H$ gate  corresponds to $\gamma{'}_{H} =\frac{\pi}{2}$, $\phi_{H}(0)=\pi$, $\phi^-_{H}=\pi$,  and $\chi_H=\frac{\pi}{4}$. For the $S$ and $T$ gates, they correspond to $\gamma{'}_{S}= \gamma{'}_{T}=\pi $,   $\phi^-_{S}=- 3\pi/4 $, $\phi^-_{T}=- 7\pi/8$, and $\phi_{S}(0)=\phi_{T}(0)=0$. Based on TOC and solving Eq. (\ref{Line}), here, $\phi(t)$  is in the form of a linear function and $\chi$ is a constant, whose path in the Bloch sphere is illustrated in Fig. \ref{fig1}(c).

\begin{figure}[tbp]
\includegraphics[width= 0.85\linewidth]{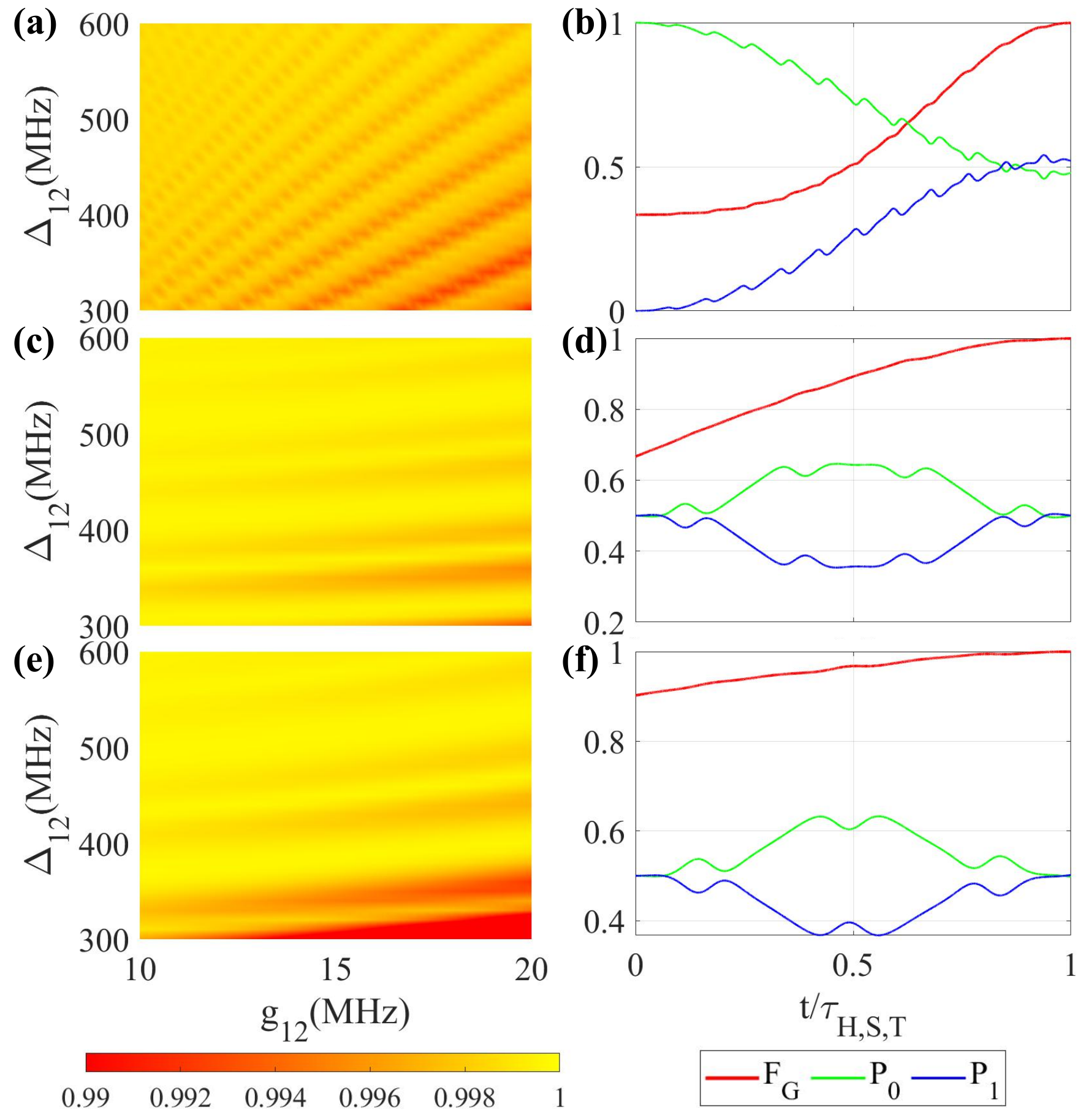}
\caption{The gate fidelity as a function of the qubits' frequency differences $\Delta_{12}$ and the coupling strength $g_{12}$. The numerical results of $H$, $S$, and $T$ gates are shown in panels (a), (c), and (e), respectively. The dynamics of the state population and the fidelity of $H$, $S$, and $T$ gates are shown in panels (b), (d), and (f), respectively. $\mathrm{F}_{\mathrm{G}}$ is the gate fidelity; $\mathrm{P}_{0}$ and $\mathrm{P}_{1}$ are the populations of the logical states $|0\rangle_{L}$ and $|1\rangle_{L}$, respectively.}
\label{fig2}
\end{figure}

\begin{figure*}[tbp]	
\includegraphics[width=1 \linewidth]{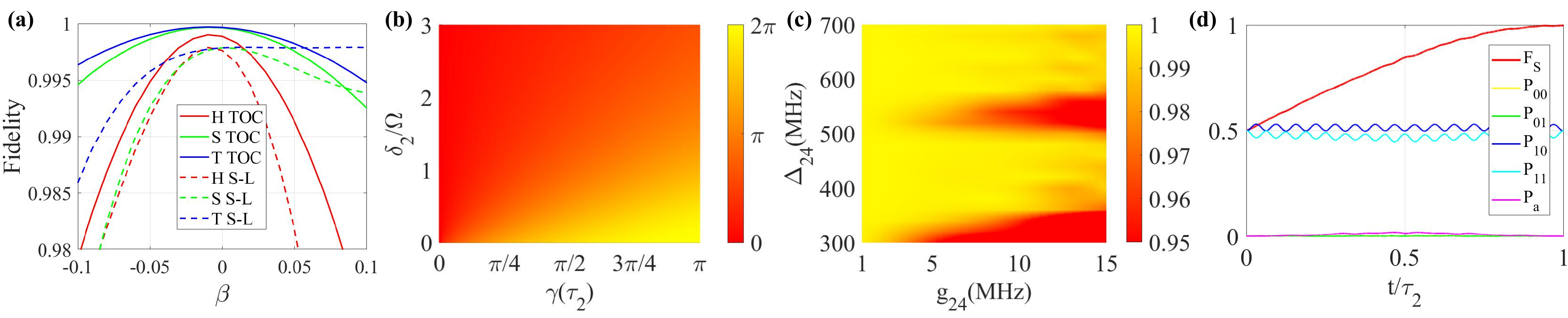}
\caption{ (a) Comparative results for the gate  robustness. Frequency drift error of TOC-based (solid line) and  S-L-based gates (dashed  lines). (b) The operation time $\tau_{2}$ in units of $1/\Omega$ with respect to the rotation angle $\gamma(\tau_{2})$ and $\delta_2/\Omega$. (c) State fidelity as the function of the qubits' frequency differences $\Delta_{24}$ and their capacitive coupling strength $g_{24}$. (d) Considering the adjacent interactions from $T_1$ and $T_3$, state population and fidelity dynamics of the CP-gate process  with prescribed  parameters as presented in the maintext, where $\mathrm{F}_{\mathrm{S}}$ is the state fidelity with the initial state $(|10\rangle_{L}+|11\rangle_{L})/\sqrt{2}$, and $\mathrm{P}_{00}$, $\mathrm{P}_{01}$, $\mathrm{P}_{10}$, $\mathrm{P}_{11}$, and $\mathrm{P}_{a}$ are the populations of $|00\rangle_{L}$, $|01\rangle_{L}$, $|10\rangle_{L}$, $|11\rangle_{L}$, and $|a\rangle$, respectively.} 
	\label{fig4}
\end{figure*}

Next, we use the master equation 
\begin{eqnarray}
\label{ZF}
\dot{\rho}=-i\left[\mathcal{H}^{'}_{12}(t), \rho\right]+
\sum_{k=1}^{N}\left(
\frac{r^-_{k}}{2}  \mathcal{A}\left(b_{k}\right)+\frac{r^{z}_k}{2} \mathcal{A}\left(b_{k}^{z}\right)\right),
\end{eqnarray}
where $b_{k}^{z}=b_{k}^{\dagger}b_{k}$, to simulate the performance of our scheme for the single-logical-qubit gates with $\mathcal{A}(b)=2b\rho b^{+}-b^{+}b\rho-\rho b^{+}b$, where $\rho$ is density operator of the quantum system with  $N=2$. And
$r_1^-=r_2^-= r^-=2\pi\times 4$ KHz and $r_1^z=r_2^z =r^z=2\pi\times 4$ KHz\cite{8} are the decay and dephasing rates of the two transmons qubits $T_1$ and $T_2$, respectively, which correspond to $\tau^-=1/r^-\approx 40$ $\mu$s and $\tau^z=1/(2 r^z) \approx 20$ $\mu$s, respectively.
As shown in Fig. \ref{fig2}, taking $g_{12}$ and $\Delta_{12}$ as variables, we  numerically obtain the fidelity of the $H$, $S$ and $T$ gates, which are defined as $F=\textrm{Tr}(U^\dagger U')/\textrm{Tr}(U^\dagger U)$, where $U'$ represents the  evolution matrix under decoherence. For typical examples, we consider  the parameters of the physical qubits as follows. The qubit frequency difference $\Delta_{12}=2\pi \times 520$ MHz, the capacitive coupling strength   $g_{12}=2\pi\times 14.5$ MHz; the detunings of the $H$, $S$, and $T$ gates are modulated to $\delta_H=2\pi\times 29.58$ MHz, $\delta_S= 2\pi\times 25$ MHz, and $\delta_T=2\pi\times 15 $ MHz; $\Gamma_2$ is set as 1.5; and  $\Omega=2\pi\times 16.18$ MHz.
With these settings, the gate operation times of the $H$, $S$, and $T$ gates are 21.9, 9.5, and 7.8 ns, respectively, and the fidelities of the $H$, $S$, and $T$ gates can reach $F_H$=99.89\%, $F_S$=99.96\%, and $F_T$=99.97\%, respectively.

Next, to test the gate robustness of our scheme, we consider the crosstalk-induced qubit-frequency drift error of the two transmon qubits $T_1$ and $T_2$, which is the main error source of the superconducting qubit lattice and is in the form of $\omega_{1, \beta}=\omega_{1}+\beta \Omega$ and $\omega_{2, \beta}=\omega_{2}-\beta \Omega$.
Under the interaction picture, the interaction Hamiltonian with error can be expressed as
\begin{eqnarray}
\mathcal{H}^{\prime}_{12,\beta} = \mathcal{H}^{\prime}_{12}+\beta \Omega\left ( b_{1}^{+}b_{1}-b_{2}^{+}b_{2} \right )
\end{eqnarray}
As shown in Fig. \ref{fig4}(a), we found that under the effect of qubit frequency drift, our scheme exhibits a better resistance than the single-loop (S-L)  scheme \cite{sl}.

\subsection{Two-logical-qubit gates via TOC}
We next consider the implementation of the controlled phase (CP) gate, which is an important element for the universal quantum gates. As shown in Fig. \ref{fig1}(a), we consider a two-logical qubits unit with two pairs of transmon qubits, $T_1$ and $T_2$, and $ T_3$ and $T_4$. Assuming $|\text{CDEF}\rangle=|\text{C}\rangle_{i}\otimes|\text{D}\rangle_{j} \otimes|\text{E}\rangle_{k}\otimes|\text{F}\rangle_{l}$, there exists a four-dimensional DFS
$S_2=\text{Span}\{|00\rangle_{L}=|1010\rangle, |01\rangle_{L}=|1001\rangle, |10\rangle_{L}=|0110\rangle,  |11\rangle_{L}=|0101\rangle\}$. In addition, an auxiliary state $|a\rangle=|0200\rangle$ is needed to assist the implementation of the CP gate. We consider the interaction between two adjacent physical qubits $T_2$ and $T_4$. Similar to the single-logical-qubit case, the frequency of the $T_2$ qubit  needs to be modulated as $\omega_2=\omega_{20}+\epsilon_2 \cos(\nu_2 t+\phi_2)$ to achieve tunable coupling between qubits $T_2$ and $T_4$.

Assuming $T_1$ and $T_3$ are in the ground state, the interacting Hamiltonian can be written as
\begin{eqnarray}\label{h42}
\mathcal{H}_{4 2}^{\prime}&=&\frac{\delta_2}{2} (|a\rangle\langle a|-|11\rangle_{L}\langle 11|) \notag\\
&+& \{g_{4 2}\mathcal{K}^{\prime}_{2}e^{i \Delta_{4 2}t}(|10\rangle_{4 2}\langle 01| +\sqrt{2}e^{i (\alpha_2+\delta_2)t}|11\rangle_{L}\langle a|
\notag\\
&+& \sqrt{2}e^{-i (\alpha_4+\frac{\delta_2}{2}) t}| 20\rangle_{4 2}\langle 11|)+\text { H.c.} \},
\end{eqnarray}
where $\mathcal{K}^{\prime}_{2}=\sum_{n  =  -\infty}^{+\infty} J_{n}\left(\Gamma_{2}'\right) \exp \left[-i n\left(\nu_{2} t+\phi_{2}\right)\right]$.
When we choose the resonance frequency $\nu_{2}=\Delta_{24}-\alpha_{2}-\delta_2$, see Eq. (\ref{HT1t}), and assume $\Omega=2g_{42}J_1(\Gamma_{2}')$, $\Gamma_{2}'=1.6$, and $\phi=\phi_2+\pi$, then the Hamiltonian in Eq. (\ref{h42}) reduces to
\begin{eqnarray}
\mathcal{H}^{\text{eff}}_{42}=\frac{1}{2}\left(\begin{array}{cc}
\delta_{2} & \Omega e^{-i \phi} \\
\Omega e^{i \phi} & -\delta_{2}
\end{array}\right),
\end{eqnarray}
where $|a\rangle$ and $|11\rangle_{L}$ form the set of orthogonal basis vectors, and $\phi_2=\eta t$ is a linear function according to the TOC solution. The evolution operator is shown in Eq. (\ref{YH1}). Setting $\gamma'=\pi$, we can obtain the evolution operator in the subspace $S_2$ as
\begin{eqnarray}
U(\tau_{2}) & = & \left(\begin{array}{cccc}
1 & 0 & 0 & 0 \\
0 & 1 & 0 & 0 \\
0 & 0 & 1 & 0 \\
0 & 0 & 0 & e^{i \gamma(\tau_{2})}
\end{array}\right) \begin{array}{c}
\end{array},
\end{eqnarray}
where $\gamma(\tau_{2})=\xi^{-}_2 +\pi$. In this way, the CP gate can be obtained. The gate time can be solved as
\begin{eqnarray}
\tau_{2}&=&\frac{2}{\Omega^2+\delta_{2}^2} \left\{\delta_{2}[\gamma(\tau_{2})-\pi] \right.\notag \\
&+& \left. \sqrt{\pi^2\delta_{2}^2  -\Omega^2 [\gamma(\tau_{2})^2-2\pi \gamma(\tau_{2})] } \right\}.
\end{eqnarray}
Similar to the $S$ and $T$ gates, the detuning $\delta_2$ can also be used to further accelerate  the gate time, as shown in Fig. \ref{fig4}(b), where we have set $\gamma(\tau_{2})= \pi /2$, $\delta_{2}=2\pi \times 27$ MHz, and $\delta_{2}/\Omega=2.3929$.

In order to properly evaluate the performance of the CP gate, with the initial state $|\psi_{in}\rangle=(|10\rangle_L+|11\rangle_L)/\sqrt{2} $, the effect of the frequency difference $\Delta_{24}$ and the coupling strength $g_{12}$ on the gate fidelity  is shown in Fig. \ref{fig4}(c). When the parameters are set as $\Delta_{24}=2\pi \times 600$ MHz, $g_{24}=2\pi \times 7$ MHz, $\alpha_2=2\pi \times 210$ MHz, and $\alpha_4=2\pi \times 230$ MHz, the fidelity of the CP gate can reach 99.88\%,  approaching 0.1\% gate infidelity. Actually, the leakage about the two adjacent qubits $T_1$ and $T_3$ should be considered as well. When we set $\Delta_{12}=\Delta_{34}=2\pi \times 900$ MHz, $\alpha_1=2\pi \times 200$ MHz, and $\alpha_3=2\pi \times 220$ MHz, the fidelity of CP gate can reach $99.72\%$. The state evolution process is shown in Fig. \ref{fig4}(d). In this case, $N=4$ is set in the master equation, Eq. (\ref{ZF}), and the rates of decay and dephasing for each transmon qubit are set as $r=r_k^-=r_k^z=2\pi\times 4$ kHz.

\section{Conclusion}

In conclusion, we propose a protocol for constructing universal quantum gates in a single-step via TOC combined with DFS encoding, and we suggest an implementation  on  superconducting circuits, consisting of transmon qubits. For $S$, $T$ and CP gates, by adjusting the detuning, the gate operations can be completed in an extremely short time, which leads to universal quantum gates approaching 0.1\% gate infidelity. Thus, our scheme provides a promising way towards the practical realization of fast quantum gates.

\acknowledgements
This work was supported by the Key-Area Research and Development Program of GuangDong Province (Grant No. 2018B030326001), the National Natural Science Foundation of China (Grant No. 12275090) and Guangdong Provincial Key Laboratory (Grant No. 2020B1212060066).

\end{document}